\documentclass[twocolumn]{aastex62}

\newcommand{\teff}{T_{\rm{eff}}}
\usepackage{xcolor}	
\usepackage{graphicx}	

\begin{document}

\title{Visual Orbits of Spectroscopic Binaries with the CHARA Array. II. The eclipsing binary HD~185912}

\author{Kathryn V. Lester}
\affil{Center for High Angular Resolution Astronomy and Department of Physics \& Astronomy, \\ Georgia State University, Atlanta, GA 30302 USA}
\email{lester@astro.gsu.edu}

\author{Douglas R. Gies}
\affil{Center for High Angular Resolution Astronomy and Department of Physics \& Astronomy, \\ Georgia State University, Atlanta, GA 30302 USA}

\author{Gail H. Schaefer}  
\affil{The CHARA Array, Mount Wilson Observatory, Mount Wilson, CA 91023 USA}

\author{Christopher D. Farrington}
\affil{The CHARA Array, Mount Wilson Observatory, Mount Wilson, CA 91023 USA}

\author{Zhao Guo}   		
\affil{Department of Astronomy \& Astrophysics, Pennsylvania State University, University Park, PA 16802 USA}

\author{Rachel A. Matson}  
\affil{NASA Ames Research Center, Moffett Field, CA 94035 USA}

\author{John D. Monnier}  
\affil{Department of Astronomy, University of Michigan, Ann Arbor, MI 48109 USA}

\author{Theo ten Brummelaar}  
\affil{The CHARA Array, Mount Wilson Observatory, Mount Wilson, CA 91023 USA}

\author{Judit Sturmann}  
\affil{The CHARA Array, Mount Wilson Observatory, Mount Wilson, CA 91023 USA}

\author{Norman Vargas}  
\affil{The CHARA Array, Mount Wilson Observatory, Mount Wilson, CA 91023 USA}

\author{Samuel A. Weiss} 
\affil{Department of Physics, Southern Connecticut State University,  New Haven, CT 06515 USA}

\begin{abstract}

We present the visual orbit of the double-lined eclipsing binary, HD~185912, from long baseline interferometry with the CHARA Array. We also obtain echelle spectra from the Apache Point observatory to update the spectroscopic orbital solution and analyze new photometry from Burggraaff et al. to model the eclipses. By combining the spectroscopic and visual orbital solutions, we find component masses of $M_1 = 1.361 \pm 0.004\ M_\odot$ and  $M_2 = 1.331 \pm 0.004\ M_\odot$, and a distance of $d = 40.75\pm 0.30$ pc from orbital parallax. From the light curve solution, we find component radii of $R_1 = 1.348\pm 0.016\ R_\odot$ and $R_2 = 1.322 \pm 0.016\ R_\odot$. By comparing these observed parameters to stellar evolution models, we find that HD~185912 is a young system near the zero age main sequence with an estimated age of 500~Myr.

\vspace{30pt}

\end{abstract}


\section{Introduction}

Eclipsing binary stars are important tools for testing models of stellar evolution and creating empirical mass-luminosity relationships, specifically when the masses and radii can be determined to within $3\%$ uncertainty \citep{torres10, eker15, moya18}. For example, empirical mass-luminosity relationships are used to determine the masses of exoplanet host stars \citep{enoch10}, and binaries with A- and F-type components are used to test the treatment of convective core overshooting in evolutionary models \citep{ct18}. However, eclipsing binaries are often close binary systems with orbital periods less than seven days, in which tidal interactions and tertiary companions can significantly affect the structure and evolution of the component stars \citep{hurley02, tokovinin06}.  In order to expand the sample of binary stars to longer orbital periods where tidal interactions are negligible, long baseline interferometry must be used to measure the visual orbit to combine with the spectroscopic orbit.  We began an observing campaign at the CHARA Array and the Apache Point Observatory (APO) to measure the visual and spectroscopic orbits of double-lined binaries (SB2) in order to measure their fundamental parameters. We presented the results for our first system, HD 224355, in \citet[][Paper I]{lester19}.

\begin{deluxetable*}{cccRRRRRR}	
\tablewidth{0pt}
\tabletypesize{\normalsize}
\tablecaption{  Radial Velocity Measurements for HD~185912\label{rvtable}    }
\tablehead{ \colhead{UT Date} & \colhead{HJD-2,400,000} & \colhead{Orbital} & \colhead{$V_{r1}$} 
& \colhead{$\sigma_1$} & \colhead{Residual}  & \colhead{$V_{r2}$} & \colhead{$\sigma_2$} & \colhead{Residual}   \\  
\colhead{} & \colhead{} & \colhead{Phase} & \colhead{(km~s$^{-1}$)} & \colhead{(km~s$^{-1}$)} 
& \colhead{(km~s$^{-1}$)} & \colhead{(km~s$^{-1}$)} & \colhead{(km~s$^{-1}$)} & \colhead{(km~s$^{-1}$)}}
\startdata
2015 Aug 30  	& 57264.6896  &   0.98  &  $  93.16$  &    0.69  &  $  0.61$  & $-127.84$  &    1.04  &  $  0.73$  \\ 
2015 Dec 01  	& 57357.5389  &   0.14  &  $ -70.32$  &    0.57  &  $ -0.66$  &  $  37.05$  &    0.85  &  $ -0.16$  \\ 
2016 Sep 14  	& 57645.7060  &   0.85  &  $  46.69$  &    0.59  &  $  0.33$  &  $ -80.74$  &    0.89  &  $  0.62$  \\ 
2016 Nov 16    	& 57708.5374  &   0.07  &  $ -42.82$  &    0.55  &  $  0.25$  &  $  10.32$  &    0.85  &  $  0.29$  \\ 
2016 Dec 15  	& 57737.5269  &   0.87  &  $  54.73$  &    0.62  &  $ -0.34$  &  $ -90.14$  &    0.94  &  $  0.12$  \\ 
2017 Feb 16  	& 57801.0539  &   0.18  &  $ -73.50$  &    0.65  &  $  0.23$  &  $  40.52$  &    1.08  &  $ -0.84$  \\ 
2017 Oct 01  	& 58027.6335  &   0.84  &  $  38.65$  &    0.60  &  $ -0.94$  &  $ -75.60$  &    0.90  &  $ -1.17$  \\ 
2018 Jan 28  	& 58147.0347  &   0.46  &  $ -48.94$  &    0.57  &  $ -0.24$  &  $  14.96$  &    0.87  &  $ -0.82$  \\ 
2018 Jun 02  	& 58271.8554  &   0.80  &  $  24.85$  &    0.54  &  $  0.30$  &  $ -59.13$  &    0.83  &  $ -0.06$  \\ 
2018 Jun 25  	& 58294.8097  &   0.80  &  $  26.72$  &    0.59  &  $  0.65$  &  $ -60.51$  &    0.91  &  $  0.11$  \\ 
2018 Sep 27  	& 58388.6182  &   0.08  &  $ -47.99$  &    0.60  &  $  0.31$  &  $  15.81$  &    0.90  &  $  0.44$  \\ 
2019 Jun 19 	& 58653.8670  &   0.80  &  $  23.32$  &    0.56  &  $ -0.01$  &  $ -57.86$  &    0.85  &  $ -0.04$  \\ 
2019 Jun 20 	& 58654.8837  &   0.93  &  $  89.71$  &    0.60  &  $ -0.90$  & $-126.12$  &    0.91  &  $  0.46$  \\ 
\enddata
\end{deluxetable*}

The next spectroscopic binary in our sample is HD~185912\footnote{V1143 Cyg, HR 7484, HIP 96620; $\alpha = 19:38:41.183$, $\delta = +54:58:25.642$, $V=5.9$ mag}, which consists of a pair of F5 V stars in a 7.6 day orbital period. The first spectroscopic solution was determined by \citet{snowden69} and updated by \citet{andersen87} and \citet{behr11}. In addition, \citet{albrecht07} presented precise radial velocities from high resolution spectra as part of their study on the spin-orbit alignment using the Rossiter-McLaughlin effect. HD~185912 is also an eclipsing binary \citep{snowden69, vanhamme84, andersen87} showing slow apsidal motion with a significant relativistic component \citep[e.g.][]{dariush05, wolf10, wilson11}. This system was included in the \citet{torres10} sample of stars with accurate fundamental parameters.  HD~185912 therefore presents a rare opportunity to test the results from interferometry against those from photometry and to provide model-independent distances from orbital parallax to test against \textit{GAIA} DR2 results \citep{stassun16, stassun18}.

We present interferometric observations and the first visual orbit for this system, as well as an updated spectroscopic and photometric analysis. In Section 2, we describe our spectroscopic observations from APO and radial velocity analysis. In Section 3, we present our interferometric observations from CHARA and the visual orbit. In Section 4, we  describe the new photometry of \citet{burggraaff18} and our light curve analysis. In Section 5, we present the resulting stellar parameters and a comparison to evolutionary models. Please note, we refer to the ``primary" as the more massive, hotter star and the ``secondary" as the less massive, cooler star. Due to the orientation of the orbit, the deeper eclipse actually occurs when the secondary star is behind the primary, so our notation is opposite that of \citet{vanhamme84} and \citet{andersen87}.

\section{Spectroscopy}\label{section:spec}

\subsection{ARCES Observations}
We observed HD~185912 thirteen times from 2015 August -- 2019 June using the ARC echelle spectrograph \citep[ARCES;][]{arces} on the APO 3.5m telescope. ARCES covers $3500 - 10500$\AA\ across 107 echelle orders at an average resolving power of $R\sim30000$. Each observation was reduced in IRAF using the standard echelle procedures, including bias subtraction, one dimensional flat fielding, wavelength calibration using ThAr lamp exposures, and correction from a barycentric to heliocentric logarithmic frame. We removed the blaze function of each echelle order using the procedure of \citet{kolbas15}.

\begin{figure*}
\centering
\epsscale{1.2}
\plotone{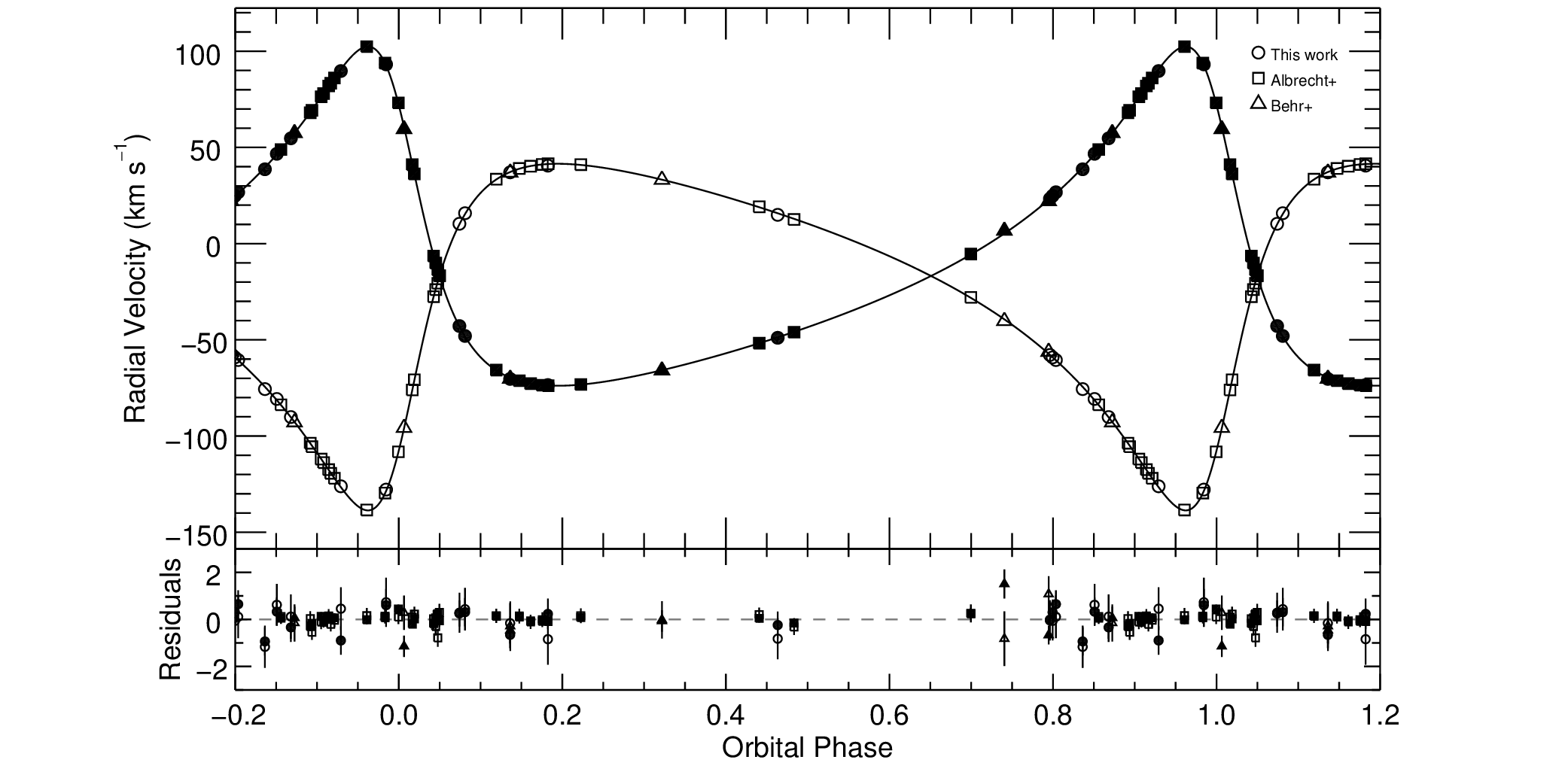}
\caption{Radial velocity curve for HD~185912 from the combined VB+SB2 solution. The filled and open points correspond to the observed velocities for the primary and secondary components, including the velocities from ARCES  (circles), \citet{albrecht07} (squares), and \cite{behr11} (triangles). The solid lines represent the model curves, and the residuals are shown in the bottom panel.  \label{rvcurve}} 
\end{figure*}

\subsection{Radial Velocities}		
We calculated the radial velocities ($V_r$) of HD~185912 using TODCOR, the two-dimensional cross correlation algorithm of \citet{todcor}, extended to multi-order spectra as described in \citet{todcor2}. Template spectra for each component were taken from BLUERED\footnote{\href{http://www.inaoep.mx/~modelos/bluered/bluered.html}{http://www.inaoep.mx/$\sim$modelos/bluered/bluered.html}} model spectra \citep{bluered} with atmospheric parameters from \citet{andersen87}. These models use solar metallicity with an abundance mixture from \citet{ag89}.     The radial velocities from each night are listed in Table \ref{rvtable}, along with the rescaled uncertainties from Section \ref{sborbit} and the residuals from the orbital solution found in Section \ref{vbsbfit}.  The monochromatic flux ratio near H$\alpha$ estimated from TODCOR is $f_2/f_1 = 0.91 \pm 0.12$.

\begin{deluxetable*}{lccc}	
\tablewidth{0pt}
\tabletypesize{\normalsize}
\tablecaption{Orbital Parameters for HD~185912 \label{orbpar}}
\tablehead{ \colhead{Parameter} & \colhead{SB2 solution}  & \colhead{VB + SB2 solution}  & \colhead{LC solution} }
\startdata		
$P$ (days)  			& $7.640735 \pm 0.000004$   	& $7.640735 \pm 0.000004$ 	& $7.640735$\tablenotemark{*}\\
$T$ (HJD-2400000) 		& $54598.1928 \pm 0.0010$  	& $54598.1930 \pm 0.0008$  	& $54598.2053 \pm0.0053$\\
$e$                    		&$  0.5380 \pm 0.0004$		& $0.5386 \pm 0.0004$		& $0.5396 \pm 0.0012$ \\
$\omega_1$ (deg)    		&$ 49.10 \pm 0.08$		& $49.11 \pm 0.10$		& $49.87 \pm 0.09$	\\
$i$ (deg)              		& \nodata 			& $86.73 \pm 0.76$		& $86.90\pm0.10$	\\
$a$ (mas)			& \nodata 			& $2.57 \pm 0.03$		& \nodata	\\
$\Omega$ (deg)         		& \nodata  			& $50.9 \pm 0.6$		& \nodata	\\   
$\gamma$ (km~s$^{-1}$)  	&$-16.81 \pm 0.04$ 		& $ -16.81 \pm 0.04$		& \nodata  \\
$K_1$ (km~s$^{-1}$)    		&$ 88.09 \pm 0.05$ 		& $  88.15 \pm 0.06$		& \nodata 	\\
$K_2$ (km~s$^{-1}$)    		&$ 90.01 \pm 0.09$ 		& $  90.08 \pm 0.08$		& \nodata 	\\ 
\enddata
\tablenotetext{*}{Fixed to spectroscopic solution.}
\end{deluxetable*}

\subsection{Spectroscopic Orbit} \label{sborbit}
We used the adaptive simulated annealing code RVFIT\footnote{\href{http://www.cefca.es/people/~riglesias/rvfit.html}{http://www.cefca.es/people/$\sim$riglesias/rvfit.html}} \citep{rvfit} to solve for the spectroscopic orbital parameters: the orbital period ($P$), epoch of periastron ($T$), eccentricity ($e$), longitude of periastron of the primary star ($\omega_1$), systemic velocity ($\gamma$), and the velocity semi-amplitudes ($K_1$, $K_2$).  We first found separate solutions for the ARCES velocities, \citet{albrecht07} velocities, and \citet{behr11} velocities,  in order to rescale the uncertainties by factors of 1.3, 1.4, and 2.4, respectively, so the reduced $\chi^2 =1$ for each dataset. Offsets of $0.1$ km~s$^{-1}$ and $-0.23$ km~s$^{-1}$ were also added to the ARCES velocities and \citet{behr11} velocities, respectively, to match $\gamma = -16.81$ km~s$^{-1}$ from \citet{albrecht07}.  Finally, we combined all data sets and refit for the spectroscopic orbital solution. The results are listed in the first column of Table \ref{orbpar}, where the uncertainties in each parameter were determined using the Monte Carlo Markov Chain feature of RVFIT. Figure \ref{rvcurve} shows the radial velocities from all data sets.

\begin{figure*}
\centering
\epsscale{0.9}
\plotone{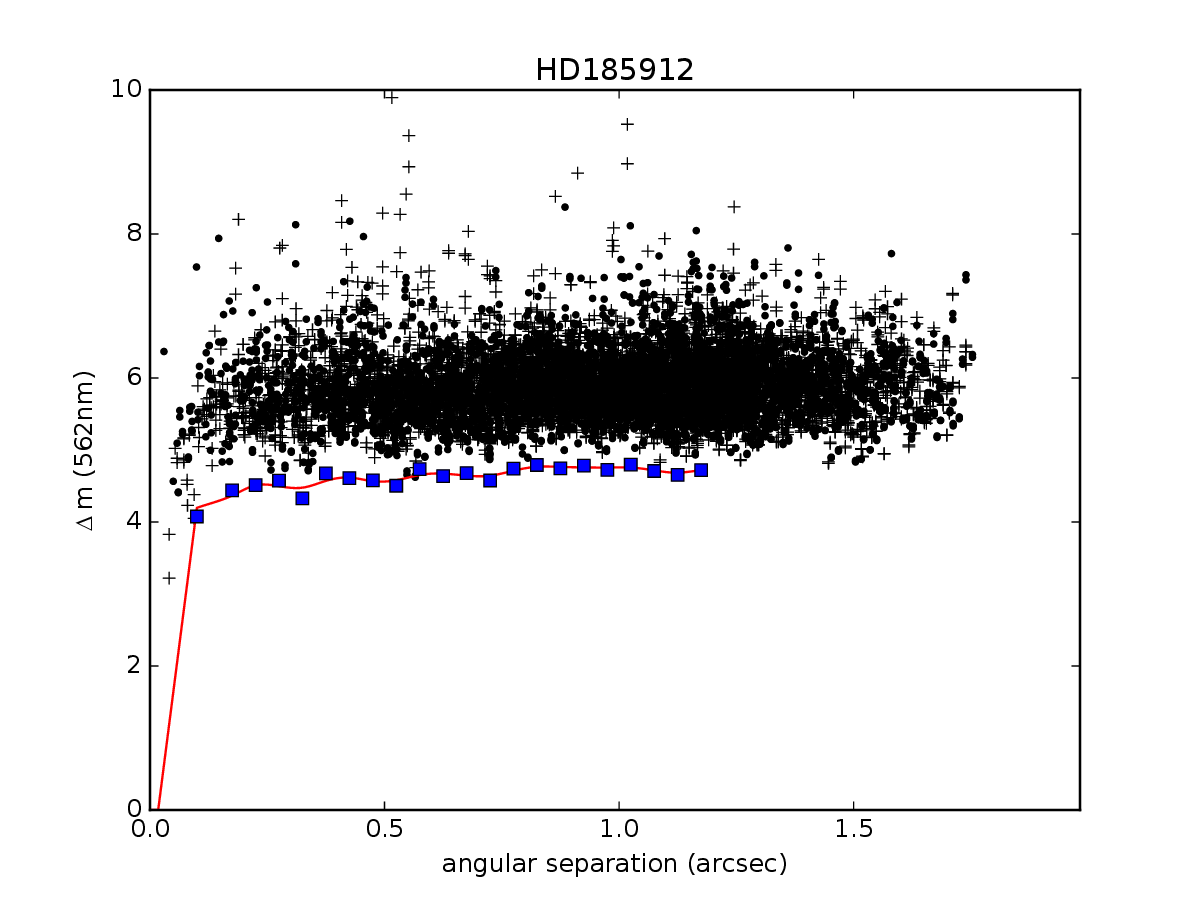}
\caption{Background sensitivity as a function of radius from the center for the reconstructed speckle image from `Alopeke. The black points represent the local maxima (crosses) and minima (dots). The blue squares mark the $5\sigma$  background sensitivity limit within $0.05\arcsec$ bins, and the red line corresponds to a spline fit. No points fall below the contrast limit, therefore no tertiary companions were detected. \label{speckle}  }   
\end{figure*}

\section{Interferometry}\label{section:inter}

\subsection{`Alopeke Observations}  
The presence of a third companion would greatly bias our results if not taken into account in our analyses, specifically affecting the resulting flux ratios, radial velocities, and orbital inclination. In order to search for the presence of a tertiary companion, HD~185912 was observed with the `Alopeke speckle imager \citep{scott18} on the Gemini North telescope\footnote{\href{https://www.gemini.edu/sciops/instruments/alopeke-zorro/}{https://www.gemini.edu/sciops/instruments/alopeke-zorro/}} in 2018 October. A set of 1000 60~ms exposures were taken in the 562~nm and 716~nm bands simultaneously and reduced using the speckle team's pipeline \citep{howell11}. Figure \ref{speckle} shows a plot of the background sensitivity limit found using the method described in \citet{horch17}. No tertiary companions were found within $1.5\arcsec$ down to a contrast of $\Delta m = 4.0$ mag.  Any more distant companions would be beyond the fields-of-view of our spectroscopic and interferometric observations.

\begin{deluxetable*}{lcclccR}	
\tablewidth{0pt}
\tabletypesize{\normalsize}
\tablecaption{CHARA/CLIMB Observing Log for HD~185912  \label{charalog}}
\tablehead{ \colhead{UT Date} & \colhead{HJD-2,400,000} & \colhead{Telescope} & \colhead{Calibrators} 
& \colhead{Number}  & \colhead{Number} & \colhead{$r_0$}  \\
\colhead{ } & \colhead{ } & \colhead{Configuration} & \colhead{ } & \colhead{of $V^2$}  & \colhead{of CP} & \colhead{(cm)} }
\startdata 	
2016 Jun 26   	&   57565.7877   &   S1-W1-E1   &   HD 178207, 187748   &   12	&   4	  & 9.8     \\
2017 May 05   	&   57878.9577   &   S2-W1-E1   &   HD 184170, 186760   &   9		&   3	  & 9.3     \\
2017 May 20   	&   57893.9517   &   S2-W1-E1   &   HD 184170, 186760   &   21	&   7	  & 9.1     \\
2017 May 21   	&   57894.9542   &   S2-W1-E1   &   HD 184170, 186760   &   21	&   7	  & 11.6   \\
2017 Aug 04   	&   57969.7870   &   S1-W1-E1   &   HD 184170, 186760   &   6		&   2	  & 10.0   \\
2017 Aug 05   	&   57970.8065   &   S1-W1-E1   &   HD 184170, 186760   &   15	&   5	  & 8.3     \\
2017 Oct 11   	&   58037.6580   &   S1-W1-E1   &   HD 184170, 186760   &   12	&   4	  & 10.6   \\
2018 Apr 10   	&   58219.0089   &   S1-W1-E1   &   HD 184170    		 &   9		&   3	  & 11.3    \\
2018 Apr 11   	&   58219.9480   &   S1-W1-E1   &   HD 184170, 186760   &   9		&   3	  & 9.5     \\
2019 Apr 26   	&   58599.9298   &   S1-W1-E1   &   HD 184170, 186760   &   21	&   7	  & 9.8     \\
2019 Apr 27   	&   58600.9608   &   S1-W1-E1   &   HD 184170, 186760   &   24	&   8	  & 8.3     \\
\enddata
\end{deluxetable*}

\begin{deluxetable*}{lccRRRRRc}
\tablewidth{0pt}
\tabletypesize{\normalsize}
\tablecaption{Relative Positions for HD~185912 \label{relpos}}
\tablehead{ \colhead{UT Date} & \colhead{HJD-2,400,000} & \colhead{Orbital} & \colhead{$\rho$} 
& \colhead{$\theta$} & \colhead{$\sigma_{maj}$} & \colhead{$\sigma_{min}$} & \colhead{$\phi$} & \colhead{$f_2/f_1$} \\
\colhead{ } & \colhead{ } & \colhead{Phase} & \colhead{(mas)} & \colhead{(deg)} &  \colhead{(mas)}  
&  \colhead{(mas)}  &  \colhead{(deg)}  &  \colhead{ }  }
\startdata 	
2016 Jun 26   	&  57565.7877 &  0.39  &   3.094 &   53.1  & 0.049 & 0.036 &   19.8  &  0.82 $\pm$  0.00  \\
2017 May 05   	&  57878.9577 &  0.38  &   3.187 & 412.8  & 0.093 & 0.059 &   17.1  &  0.96 $\pm$  0.16  \\
2017 May 20   	&  57893.9517 &  0.34  &   3.226 &   52.5  & 0.109 & 0.053 &   46.6  &  0.93 $\pm$  0.09  \\
2017 May 21   	&  57894.9542 &  0.47  &   2.787 &   53.8  & 0.031 & 0.018 &  131.1 &  0.96 $\pm$  0.01  \\
2017 Aug 04   	&  57969.7870 &  0.27  &   3.243 & 412.6  & 0.128 & 0.057 &  103.1 &  0.93 $\pm$  0.34  \\
2017 Aug 05   	&  57970.8065 &  0.40  &   3.125 & 413.9  & 0.054 & 0.036 &   63.1  &  0.99 $\pm$  0.03  \\
2017 Oct 11   	&  58037.6580 &  0.15  &   2.415 &   50.8  & 0.086 & 0.045 &   94.5  &  0.94 $\pm$  0.08  \\
2018 Apr 10   	&  58219.0089 &  0.88  &   1.212 & 225.5  & 0.070 & 0.040 &   64.3  &  0.98 $\pm$  0.02  \\
2018 Apr 11   	&  58219.9480 &  0.01  &   0.657 & 235.3  & 0.113 & 0.078 &  139.5 &  0.92 $\pm$  0.11  \\
2019 Apr 26   	&  58599.9298 &  0.74  &   0.415 & 431.3  & 0.083 & 0.076 &  175.1 &  0.83 $\pm$  0.18  \\
2019 Apr 27   	&  58600.9608 &  0.87  &   1.091 & 225.9  & 0.094 & 0.063 &  137.2 &  0.98 $\pm$  0.02  \\
\enddata  
\end{deluxetable*}

\subsection{CLIMB Observations}  
We observed HD~185912 with the CHARA Array \citep{chara} eleven times from 2016 June -- 2019 April, using the CLIMB \citep{climb} beam combiner to combine the $K'$-band light from three telescopes.  Table \ref{charalog} lists the observation dates, the telescopes and calibrator stars used, the number of data points measured, and the average Fried parameter ($r_0$) for each night. Our data were reduced with the pipeline developed by J. D. Monnier, using the general method described in \citet{monnier11} and extended to three beams \citep[e.g., ][]{kluska18}, resulting in squared visibilities ($V^2$) for each baseline and closure phases (CP) for each closed triangle. Instrumental and atmospheric effects on the observed visibilities were measured using observations of stars with known angular diameters (HD 178207, 184170, 186760 and 187748) taken before and after the target.  One calibrator-target-calibrator sequence is referred to as a ``bracket". The respective $K'$-band angular diameters from SearchCal\footnote{\href{http://www.jmmc.fr/searchcal}{http://www.jmmc.fr/searchcal}}  are $0.260 \pm 0.007$ mas, $0.592 \pm 0.014$ mas, $0.445 \pm 0.011$ mas, and $0.374 \pm 0.009$ mas \citep{searchcal}.

\begin{figure*}	 
\centering
\epsscale{1.2}
\plotone{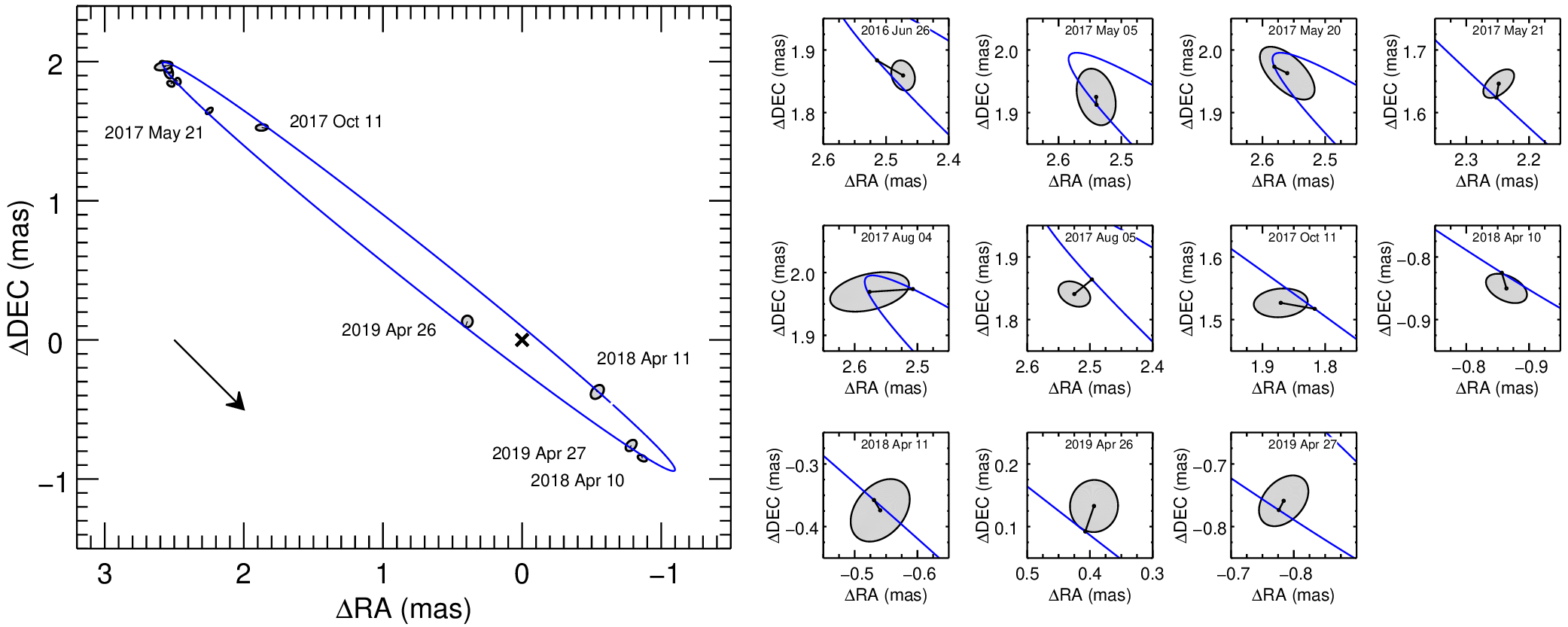}
\caption{Left: Visual orbit for HD~185912 from the combined VB+SB2 solution. The primary star is located at the origin (black cross). The relative positions of the secondary are plotted as the filled ovals corresponding to the sizes of the error ellipses, along with a line connecting the observed and predicted positions. The solid blue line shows the full model orbit, and the arrow shows the direction of orbital motion. Right: Plots of individual data points in chronological order and on the same $0.2 \times 0.2$ mas scale. \label{vb}   }  
\end{figure*}

\subsection{Binary Positions}  
Binary positions were measured using the grid search code\footnote{\href{http://chara.gsu.edu/analysis-software/binary-grid-search}{http://chara.gsu.edu/analysis-software/binary-grid-search}} of \citet{schaefer16}. We estimated the angular diameters of both components to be $0.26$ mas using the \textit{GAIA} DR2 parallax \citep{gaia1, gaia2} and the radii from \citet{andersen87}.  Both stars are smaller than the $0.6$ mas angular resolution of CLIMB and therefore unresolved, so we held the angular diameters fixed and fit only for the relative position of the secondary component and the flux ratio,  as described in Paper I.  Table \ref{relpos} lists the separation and position angle of the secondary component (measured east of north) for each night, the major axis, minor axis and position angle of the error ellipse, and the best-fit flux ratio at 2.13$\mu$m. The weighted average flux ratio from all nights is $f_2/f_1 = 0.97 \pm 0.06$. The sizes of the error ellipses depend on several factors, including the number of brackets obtained, the telescope combination used, the seeing, and the data quality.  Figure \ref{vb} shows the observed relative positions, as well as the best-fit visual orbit found in the next section.

\subsection{Combined Visual + Spectroscopic Solution}\label{vbsbfit}
From the visual orbit alone, one can determine the orbital inclination ($i$),  angular semi-major axis ($a$), and longitude of the ascending node ($\Omega$).  By combining the interferometric and spectroscopic data, we can fit for all ten orbital parameters ($P$, $T$, $e$, $i$, $a$, $\omega_1$, $\Omega$, $\gamma$, $K_1$, $K_2$) using the method of \citet{schaefer16}, described in detail in Paper I. The best fit orbital parameters for this combined (VB+SB2) solution are listed in the third column of Table \ref{orbpar}, along with the uncertainties calculated using a Monte Carlo error analysis.  The best-fit model radial velocity curves are shown in Figure \ref{rvcurve} and model visual orbit is shown in Figure \ref{vb}.

\break

\section{Photometry}\label{section:phot}

\subsection{MASCARA Light Curve}  
\begin{figure*}	 
\centering
\epsscale{1.2} 
\plotone{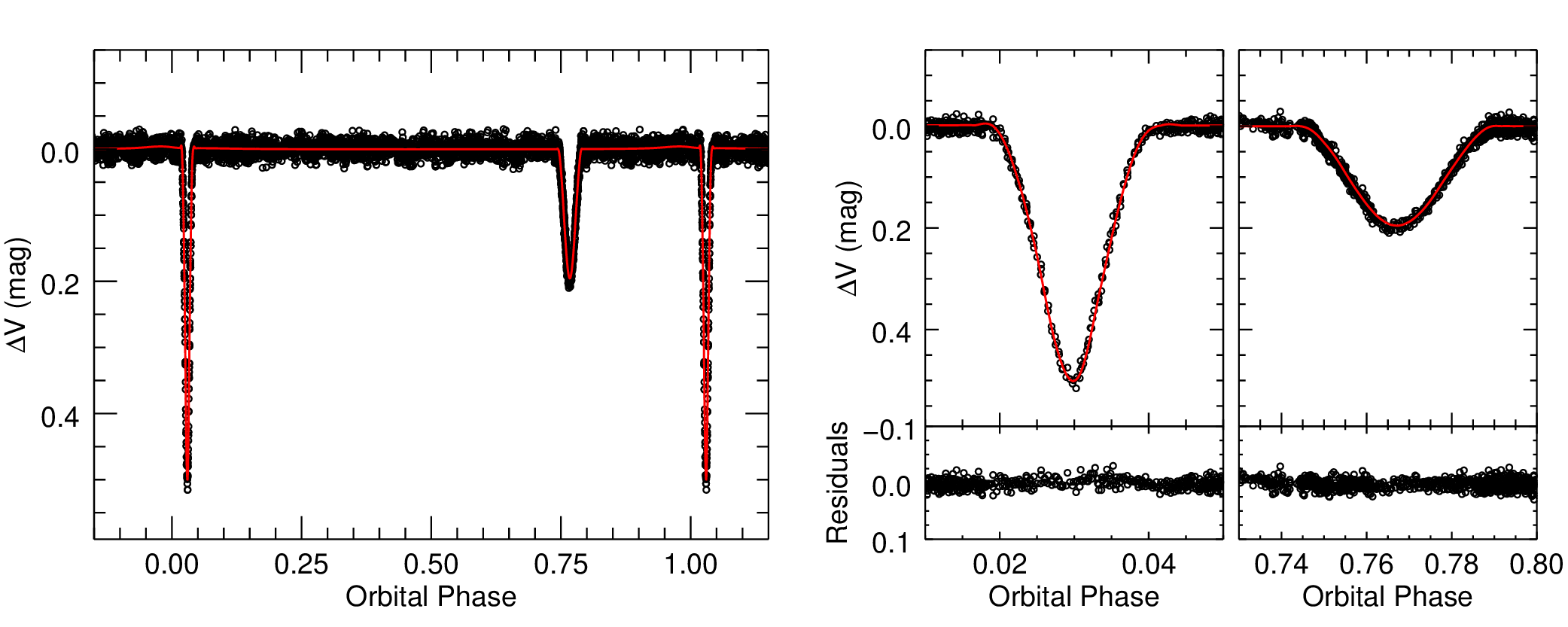}
\caption{  
Light curve of HD~185912 from \citet{burggraaff18} photometry. The full, phased light curve is shown in the left panel, with detailed views of the primary and secondary eclipses in the right panels. Phase 0 corresponds to the time of periastron. The best fit ELC model is shown as the solid red line, which was used to calculated the residuals shown in the righthand panels. \label{lc}   }  
\end{figure*}

HD~185912 was recently observed by the Multi-site All-Sky CAmeRA (MASCARA\footnote{\href{http://mascara1.strw.leidenuniv.nl/}{http://mascara1.strw.leidenuniv.nl/}}) photometric survey of \citet{burggraaff18}, who completed $V$-band relative photometry of bright stars in search of exoplanets.  The observations spanned ten orbital cycles, but the primary and secondary eclipses were observed fully in only two. We first removed the systematic effects as a function of lunar phase and sidereal time as described in their paper and folded the data using the orbital period from the spectroscopic solution. We then removed outlier points by calculating the residuals against a model light curve with parameters from \citet{andersen87} and discarding all of the points outside three times the standard deviation. The folded light curve is shown in Figure \ref{lc}.  

\subsection{Light Curve Modeling}\label{lcfit}
We modeled the light curve using the Eclipsing Light Curve code of \citet{elc}. We held the orbital period fixed to the spectroscopic solution and used ELC's genetic optimizer to fit for $T$, $e$, $i$, and $\omega_1$, as well as the relative radius of each component ($R_1/a$, $R_2/a$) and the temperature ratio ($T_{\rm{eff}\ 2}/T_{\rm{eff}\ 1}$). We found that $T$, $e$, and $\omega_1$ were well constrained by the optimizer and are listed in Table \ref{orbpar}.  The inclination, relative radii, and temperature ratio were not well constrained, because it is difficult to determine the individual radii directly from the light curve in partially eclipsing systems with very similar components. There exists a family of solutions that fit the observations equally well, so that only the value of $(R_1 + R_2)/a$ can be determined accurately. 

\begin{figure*}
\centering
\epsscale{1.2}
\plotone{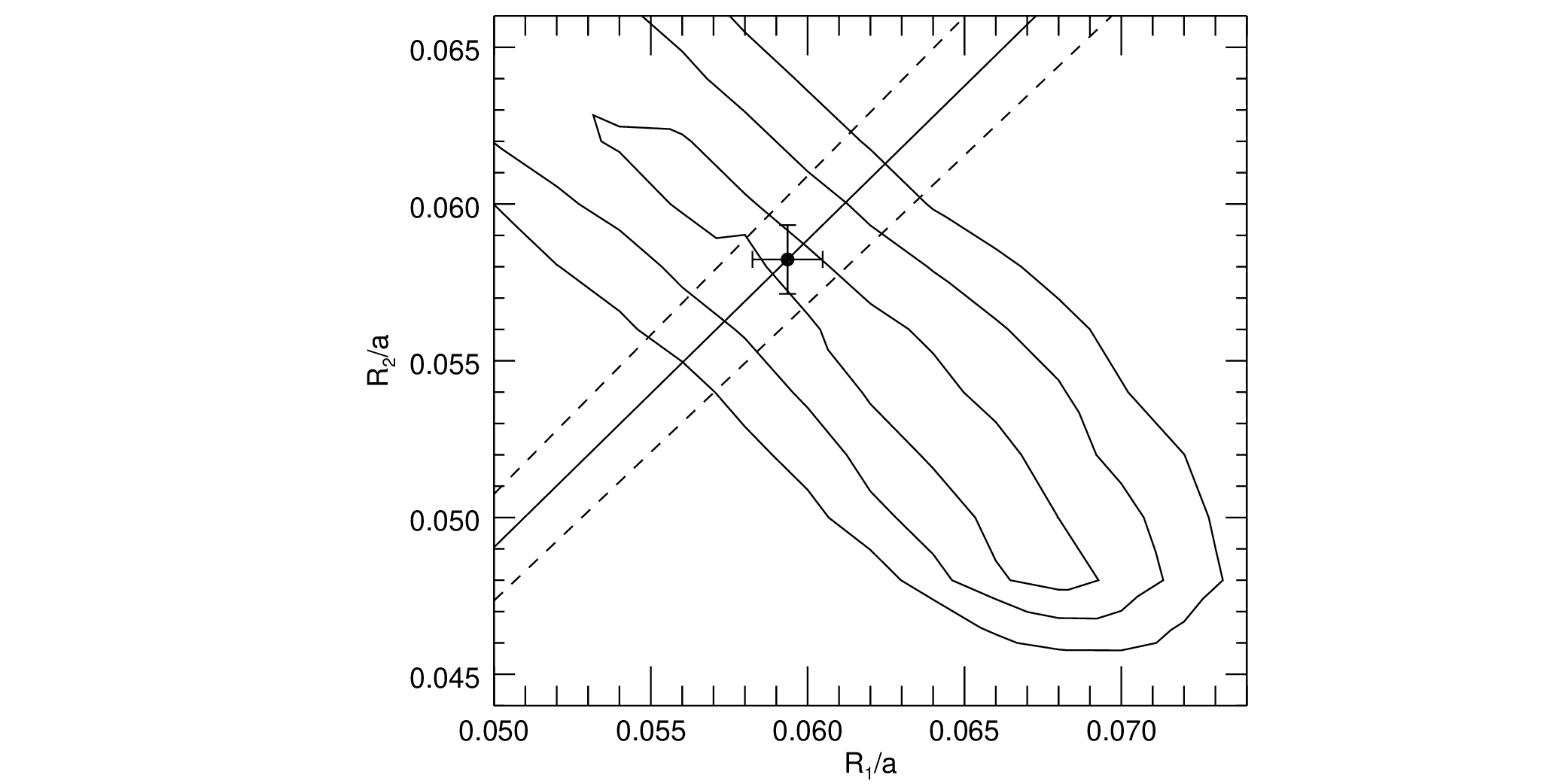}
\caption{Contour plot of $\chi^2$ as a function of relative radius (R/a), corresponding to the $1-$, $2-$, and $3-\sigma$ levels. The solid line corresponds to the mean radius ratio from Section \ref{sedfit} and the dashed lines correspond to the uncertainty. The best-fit pair of relative radii is marked with the black point.  \label{contour}} 
\end{figure*}

To show this more clearly, we calculated the $\chi^2$ goodness-of-fit statistic across the primary and secondary eclipses for model light curves over a grid of $R_1/a$ and $R_2/a$ values, fitting for the inclination and temperature ratio at each grid point. In order to weight equally the primary and secondary eclipses, we divided the $\chi^2$ values for each eclipse by the number of points within each eclipse (124 and 458) before adding the $\chi^2$ values together. Figure \ref{contour} shows the $\chi^2$ contour as a function of relative radius, where the valley of possible solutions is easily visible.

Solving the problem of partially eclipsing systems therefore requires additional constraints; for example, \citet{andersen87} used the luminosity ratio from their spectroscopic analysis to inform their results. We used the observed flux ratios and model surface fluxes to estimate a radius ratio (see Section \ref{sedfit}), plotted as the solid line in Figure \ref{contour}. We found the minimum $\chi^2$ value along this line to correspond to 
$i= 86.9 \pm 0.1$ deg, 
$R_1/a = 0.0594 \pm 0.0011$, 
$R_2/a = 0.0582 \pm 0.0011$, and 
$T_{\rm{eff}\ 2}/T_{\rm{eff}\ 1} = 0.99 \pm 0.01$. 
The uncertainties correspond to where $\chi^2 \le \chi^2_{min} + 1$.  This inclination is consistent with that from the visual orbit, however this value does depend on the relative radii and surface flux models while the visual orbit is independent of models.

\section{Stellar Parameters}\label{section:param}

\subsection{Masses and Distance}
By combining the results from spectroscopy with those of interferometry, we found the component masses of HD~185912 to be $M_1 = 1.361 \pm 0.004 M_\odot$ and $M_2 = 1.332 \pm 0.004 M_\odot$. By combining the angular and physical sizes of the orbit, we found the distance to be $d = 41.02\pm 0.22$ pc. This is consistent with the \textit{Hipparcos} distance of $d = 40.88 \pm 0.48$ pc \citep{hipparcos1, hipparcos2} and the \textit{GAIA} DR2 distance of $d = 40.47 \pm 0.08$ pc \citep{gaia1, gaia2}.

\begin{deluxetable}{lcc} 
\tablewidth{0pt}
\tabletypesize{\normalsize}
\tablecaption{Stellar Parameters of HD~185912 \label{atmospar}}
\tablehead{ \colhead{Parameter} & \colhead{Primary} & \colhead{Secondary} }
\startdata		
Mass ($M_\odot$)			& $1.361 \pm 0.004$		& $1.332 \pm 0.004$\\ 
Radius ($R_\odot$)			& $1.348\pm 0.016$		& $1.322 \pm 0.016$	\\ 
$\teff$ (K)					& $6620 \pm 190$ 		& $6570 \pm 220$ 	\\
Luminosity ($L_\odot$)		& $ 3.35\pm0.44$		& $3.13\pm0.50$	\\
$\log g$ (cgs)				&  $4.31 \pm 0.03$ 		& $4.32 \pm 0.04$  	\\ 
$V \sin i$ (km~s$^{-1}$)		& $19.1 \pm 0.6$ 		& $27.9 \pm 1.2$	\\ 
Semi-major axis ($R_\odot$)  	& \multicolumn{2}{c}{$22.71 \pm 0.03$} 		\\ 
Distance (pc)				& \multicolumn{2}{c}{$41.02\pm 0.22$ } 		\\ 
$E(B-V)$ (mag)				&\multicolumn{2}{c}{$0.08 \pm 0.01$}		\\   
\enddata	
\end{deluxetable}	

\subsection{Effective Temperatures and Rotational Velocities}\label{tempfit}
We first used the Doppler tomography algorithm of \citet{tomography} to reconstruct the individual spectrum of each component for all echelle orders between $4000-7000$\AA. We then cross-correlated the reconstructed spectra with BLUERED models of different effective temperatures to find the best-fit temperature for each echelle order. The maximum correlation for each order was used to calculate the weighted average temperature for each component,  where better correlated orders were more highly weighted,  and the uncertainty corresponding to the standard deviation of the temperatures from all orders. We found the effective temperatures to be $T_{\rm{eff}\ 1} = 6620 \pm 190$~K and $T_{\rm{eff}\ 2} = 6570 \pm 220$~K.  

These values are higher than those determined by \citet{smalley02} from the Balmer line profiles ($T_{\rm{eff}\ 1} = 6441 \pm 201$~K and $T_{\rm{eff}\ 2} = 6393 \pm 136$~K), but consistent with the values determined by \citet{wilson11} from absolute photometry ($T_{\rm{eff}\ 1} = 6653 \pm 11$~K and $T_{\rm{eff}\ 2} = 6558 \pm 5$~K). However, the latter uncertainties are rather underestimated; the authors included internal uncertainties from the least squares fitting procedure and calibration of the filter passbands in their code, but did not incorporate uncertainties in the observations from comparison star magnitudes nor uncertainties in the fixed model parameters.

We used a similar method as described above to determine the projected rotational velocity ($V \sin i $) of each component by cross-correlating model spectra of different $V \sin i $ with the reconstructed spectra. We found $V_1 \sin i = 19.1 \pm 0.6$~km~s$^{-1}$ and $V_2 \sin i = 27.9 \pm 1.2$~km~s$^{-1}$.  These rotational velocities are consistent with the more precise values found by \citet{albrecht07} ($19.6 \pm 0.1$~km~s$^{-1}$ and $28.2 \pm 0.1$~km~s$^{-1}$). Both components are also rotating slower than the projected pseudo-synchronous velocities of $31.1$~km~s$^{-1}$ and $30.5$~km~s$^{-1}$.

\begin{figure*}
\centering
\epsscale{1.1}
\plotone{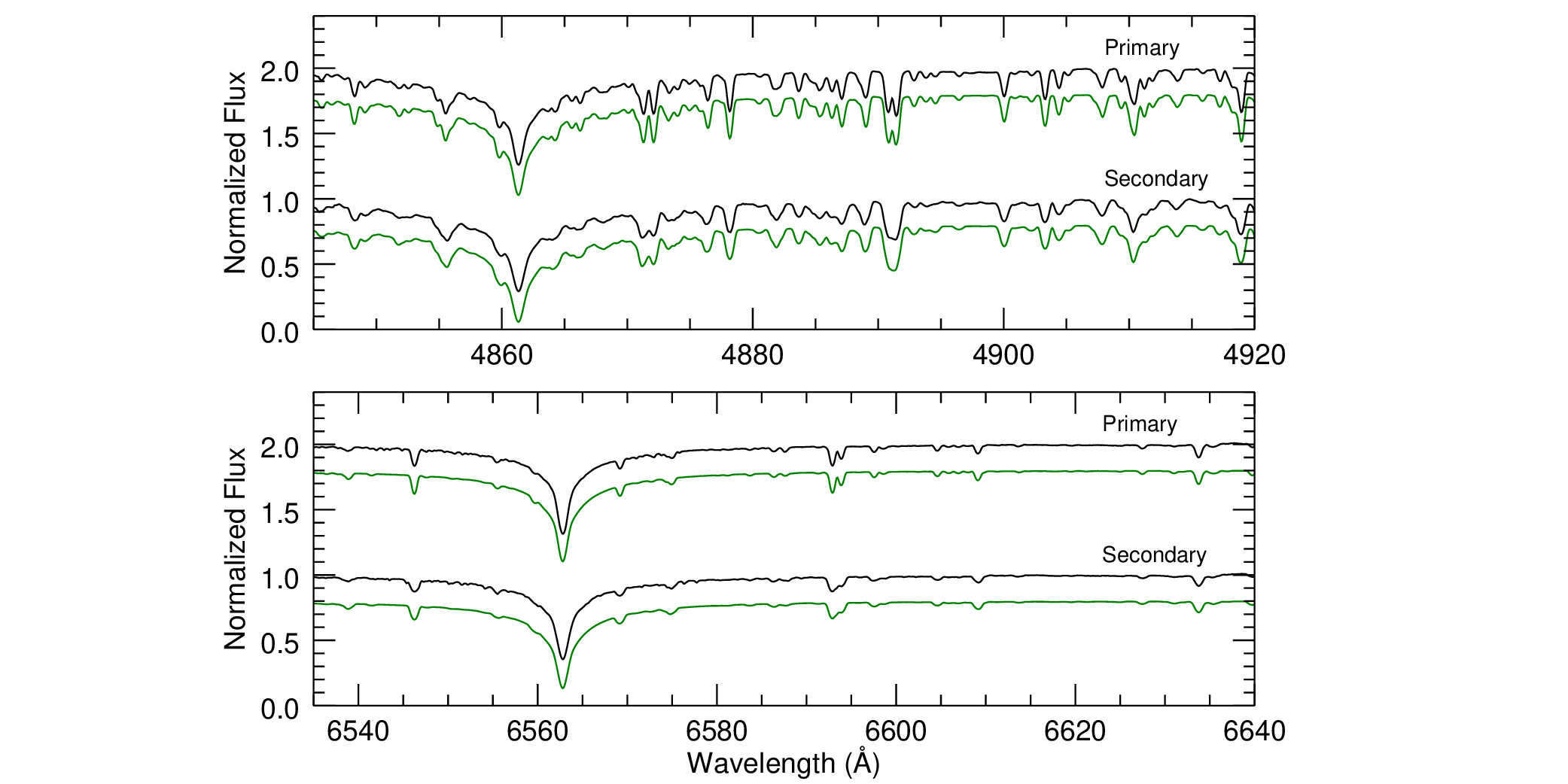}
\caption{Reconstructed spectra of HD~185912 (black) and best-fit model spectra (green) near H$\alpha$ and H$\beta$ using the atmospheric parameters in Table~\ref{atmospar}. The model spectra are offset by $-0.2$ for clarity. \label{balmer}} 
\end{figure*}

\subsection{Radii and Surface Gravities}\label{sedfit}	
We created surface flux models of each component from ATLAS9 model atmospheres \citep{sedmodel} using the temperatures found in the previous section. By comparing the observed flux ratios and model surface fluxes, we calculated the radius ratio to be $R_2/R_1 = 0.96 \pm 0.08$ near H$\alpha$ from the spectroscopic flux ratio and $R_2/R_1 = 0.99 \pm 0.04$ in $K'$-band from the interferometric flux ratio.  The weighted average radius ratio is $R_2/R_1 = 0.98 \pm 0.04$. We then used this radius ratio to determine the individual stellar radii from two methods; spectral energy distribution (SED) fitting and light curve fitting.  

\begin{figure*}
\centering
\epsscale{1.1}
\plotone{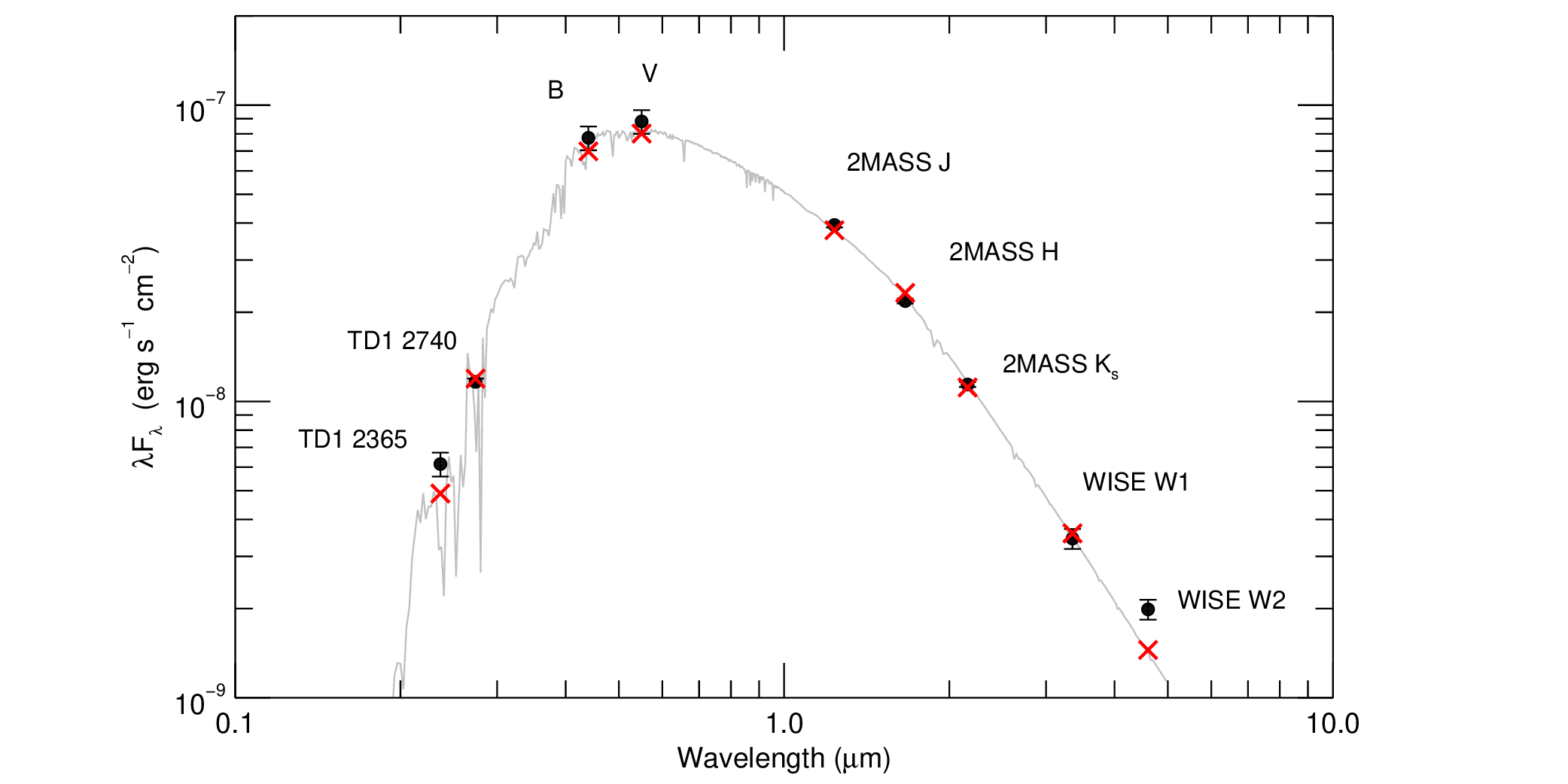}
\caption{Observed photometry of HD~185912 (black filled circles) and best-fit model fluxes (red crosses). A full binary model SED is shown in grey for reference. \label{sed}} 
\end{figure*}

For the first method, we took broad-band photometry from the literature to create the SED for HD~185912 shown in Figure \ref{sed}, which includes ultraviolet data from TD1 \citep{TD1}, optical data from \citet{ubv}, and infrared data from 2MASS \citep{2mass} and WISE \citep{wise}. We then created a binary SED model to compare to the observed SED by integrating the surface flux models across each photometric passband, and then fit for the primary angular diameter and reddening (see Section 5.2 of Paper I). We found angular diameters of $\theta_1 = 0.32\pm0.01$ mas and $\theta_2 = 0.31\pm0.01$ mas, which correspond to stellar radii of  $R_1 = 1.39\pm 0.04 R_\odot$, $R_2 = 1.37 \pm 0.06 R_\odot$, and a reddening value of $E(B-V)=0.08 \pm 0.01$ mag.  

For the second method, we used the orbital parameters found in Section~\ref{vbsbfit} and the relative radii found in Section~\ref{lcfit} to calculate the individual stellar radii. We found $R_1 = 1.348 \pm 0.016 R_\odot$ and $R_2 = 1.322 \pm 0.016 R_\odot$, corresponding to surface gravities of $\log g_1 = 4.31 \pm 0.03$ and $\log g_2 = 4.32 \pm 0.04$  as listed in Table~\ref{atmospar}. Both methods provide consistent results, but this is expected since they depend on the same model fluxes and radius ratio.  Using these radii from the light curve solution and the effective temperatures, we calculated the luminosities of each component to be $L_1 = 3.35\pm0.44 \ L_\odot$ and $L_2 = 3.13\pm0.50 \ L_\odot$ from the Stefan-Boltzmann law.

\subsection{Comparison with Evolutionary Models} 
\begin{figure*}
\centering
\epsscale{1.1}
\plotone{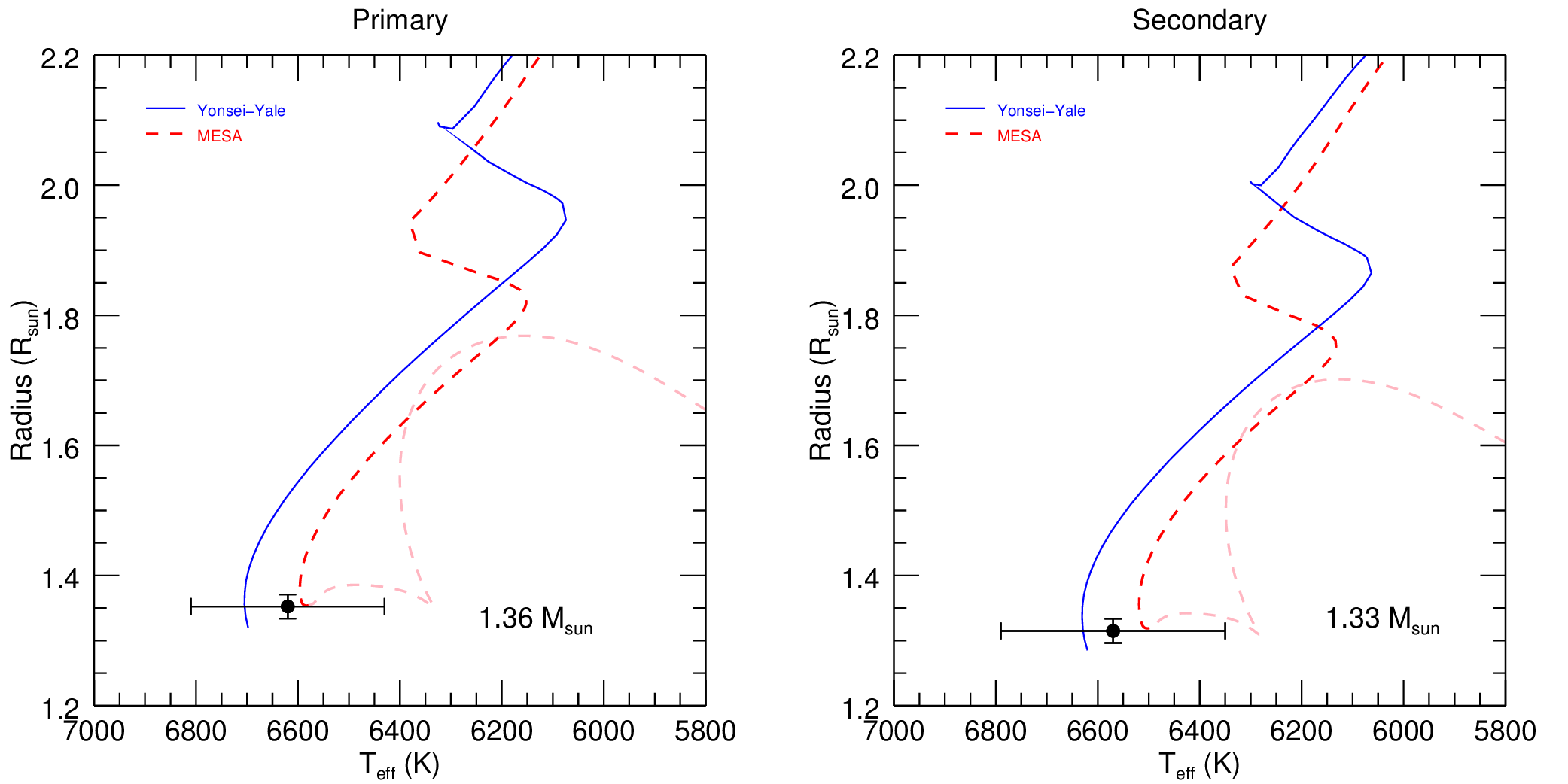}
\caption{ Evolutionary tracks for each component of HD~185912. The solid blue lines are the Yonsei-Yale models. The dashed red lines are the MESA models, with the pre-main sequence portion shown in light red. \label{evo}} 
\end{figure*}

We created model evolutionary tracks for each component of HD~185912 using the Yonsei-Yale (Y$^2$) evolutionary models of \citet{y2} and the MESA stellar evolution code of \citet{mesa1, mesa2, mesa3, mesa4, mesa5}, shown in Figure~\ref{evo}. The Yonsei-Yale models\footnote{\href{http://www.astro.yale.edu/demarque/yystar.html}{http://www.astro.yale.edu/demarque/yystar.html}} were created using the model interpolation program, and the MESA models\footnote{\href{http://www.mesa.sourceforge.net}{http://www.mesa.sourceforge.net}} were created using MESA release 10108 with overshooting parameters for each component taken from the empirical relationship of \citet{ct18}. Both sets of models are non-rotating and use solar metallicity. The Yonsei-Yale models use the solar abundance mixture from \citet{g96}, while the MESA models use the mixture of \citet{gs98}.

As seen in Figure~\ref{evo}, HD~185912 lies very close to the zero age main sequence.   We estimated the age of each component based on the portions of the evolutionary tracks that lie within the observed uncertainties, then took the average to estimate system ages of 550 Myr from the Yonsei-Yale models and 100 Myr from the MESA models. The individual ages of each component from their evolutionary tracks are consistent to within 5\%. This young age is confirmed by the presence of the \ion{Li}{1} $6708$\AA\ absorption line in our spectra.

\section{Discussion}
We determined the first visual orbit for HD~185912 from long baseline interferometry with the CHARA Array, as well as  updated spectroscopic orbits and photometric analysis. From the combined visual and spectroscopic solution, we found the component masses to within 0.3\% and the distance to within 0.8\%. We found the component radii to within 5\% from SED fitting and to within 1.2\% from light curve modeling, but these errors are likely underestimated in partially eclipsing systems. Therefore, more precise photometry during the eclipses is needed to determine the individual radii, such as the highly anticipated \textit{TESS} observations currently underway in the northern hemisphere \citep{tess}.

By comparing our observed stellar parameters to evolutionary models, we found that HD~185912 is a young system located on the zero age main sequence and likely in the process of tidal circularization \citep{meibom05}. We checked for membership in 29 nearby moving groups using the BANYAN\footnote{\href{http://www.exoplanetes.umontreal.ca/banyan/banyansigma.php}{http://www.exoplanetes.umontreal.ca/banyan/banyansigma.php}} website \citep{banyan}, which compares the position, proper motion, radial velocity, and parallax to that of each moving group. BANYAN reported a membership probability of $0\%$ for all associations, so HD~185912 is simply a young field star.

Eclipsing binaries like HD~185912 are important for comparing the results from interferometry and photometry. Specifically, the orbital inclination from interferometry is consistent with the results from photometry, providing a proof of concept for our project. We are continuing interferometric observations of several other longer period spectroscopic binaries to determine their visual orbits and determine their fundamental stellar parameters.

\acknowledgments

{\footnotesize{
The authors would like to thank the staff at APO and CHARA for their invaluable support during observations, as well as the anonymous referee for their insightful comments. Institutional support has been provided from the GSU College of Arts and Sciences and the GSU Office of the Vice President for Research and Economic Development. This work is based in part upon observations obtained with the Georgia State University Center for High Angular Resolution Astronomy Array at Mount Wilson Observatory, supported by the National Science Foundation under Grants No. AST-1636624 and AST-1715788; the Apache Point Observatory 3.5-meter telescope, owned and operated by the Astrophysical Research Consortium; and the High-Resolution Imaging instrument `Alopeke at Gemini-North (GN-2018B-FT-102), funded by the NASA Exoplanet Exploration Program and built at the NASA Ames Research Center by Steve B. Howell, Nic Scott, Elliott P. Horch, and Emmett Quigley. The Gemini Observatory is operated by the Association of Universities for Research in Astronomy, Inc. This work has also made use of the Jean-Marie Mariotti Center SearchCal service, the CDS Astronomical Databases SIMBAD and VIZIER, data from the Wide-field Infrared Survey Explorer, and data from the Two Micron All Sky Survey. }}

\facilities{CHARA, APO:3.5m, Gemini:North}


\end{document}